# Pseudo 5D HN(C)N Experiment to Facilitate the Assignment of Backbone Resonances in Proteins Exhibiting High Backbone Shift Degeneracy

Dinesh Kumar,[1,#] Nisha Raikwal,[1] Vaibhav Kumar Shukla,[2] Himanshu Pandey,[2] Ashish Arora[2] and Anupam Guleria[1]

[1]*Centre of Biomedical Research (CBMR), SGPGIMS Campus, Raibareli Road, Lucknow-226014, India*
[2]*Molecular and Structural Biology Division, CSIR, Central Drug Research Institute, Lucknow 226 001, India*

[#]**Author for Correspondence:**

Dr. Dinesh Kumar
(Assistant Professor)
Centre of Biomedical Research (CBMR),
SGPGIMS Campus,
RB Road, Lucknow-226014
Uttar Pradesh, India
Mobile: +91-9044951791, +91-8953261506
Fax: +91-522-2668215
Email: dineshcbmr@gmail.com
Webpage: www.cbmr.res.in/dinesh.html

## Abstract:

Assignment of protein backbone resonances is most routinely carried out using triple resonance three dimensional NMR experiments involving amide $^1$H and $^{15}$N resonances. However for intrinsically unstructured proteins, alpha-helical proteins or proteins containing several disordered fragments, the assignment becomes problematic because of high degree of backbone shift degeneracy. In this backdrop, a novel reduced dimensionality (RD) experiment –(5,3)D-h<u>NCO</u>-<u>CAN</u>H- is presented to facilitate (and/or to validate) the sequential backbone resonance assignment in such proteins. The proposed 3D NMR experiment makes use of the modulated amide $^{15}$N chemical shifts (resulting from the joint sampling along both its indirect dimensions) to resolve the ambiguity involved in connecting the neighboring amide resonances (i.e. $H_iN_i$ and $H_{i-1}N_{i-1}$) for overlapping amide NH peaks. The experiment -encoding 5D spectral information- leads to a conventional 3D spectrum with significantly reduced spectral crowding and complexity. The improvisation is based on the fact that the linear combinations of intra-residue and inter-residue backbone chemical shifts along both the co-evolved indirect dimensions span a wider spectral range and produce better peak dispersion than the individual shifts themselves. Taken together, the experiment -in combination with routine triple resonance 3D NMR experiments involving backbone amide ($^1$H and $^{15}$N) and carbon ($^{13}$C$^\alpha$ and $^{13}$C') chemical shifts- will serve as a powerful complementary tool to achieve the nearly complete assignment of protein backbone resonances in a time efficient manner. The performance of the experiment and application of the method have been demonstrated here using a 15.4 kDa size folded protein and a 12 kDa size unfolded protein.

**Introduction:**

Over the decades, NMR has emerged as a powerful technique for studying the structure and dynamics of proteins and their complexes in solution. Further, it is the technique of choice for studying conformational properties of intrinsically unstructured proteins (IDPs), and their interactions with their physiological binding partners in solution [1-3]. For various such studies on proteins by NMR, the very first and key requirement is the sequence specific assignment of backbone ($^1$H, $^{15}$N, $^{13}$C' and $^{13}$C') resonances [4,5]. The well-established and most routinely used assignment strategies involve the use of $^{15}$N, $^1$H$^N$ resolved triple resonance experiments sequentially linking $^{13}$C$^{\alpha/\beta}$, $^{13}$C' or $^{15}$N shifts [6-22] and many proteins have been assigned this way (evident from the Biological Magnetic Resonance Bank: http://www.bmrb.wisc.edu). However for proteins exhibiting high degree of backbone amide and carbon shift degeneracy (e.g. $\alpha$-helical proteins or proteins containing disordered fragments including IDPs), getting this information in an unambiguous and time-efficient manner has always remained problematic and challenging. Therefore new or alternative NMR methods and strategies -for rapid and efficient assignment of backbone resonances in such proteins- are required.

Several efforts have been made in the past to resolve this problem [8,14,16,18,22-35] involving either (a) the use of H$^\alpha$-detected NMR experiments [29,36] or (b) $^{13}$C detected NMR experiments [24,32,37], or (c) higher dimensionality (4D or 5D) NMR experiments [30,32,34,35]. Of them, H$^\alpha$-detected experiments require protein samples in deuterated solvents, while $^{13}$C-detected experiments suffer from reduced sensitivity due to the lower gyromagnetic ratio of $^{13}$C with respect to $^1$H [18]. Further if the assignment process has already been started acquiring the conventional $^{15}$N, $^1$H$^N$ resolved triple resonance 3D NMR experiments, one has to acquire the completely different set of NMR experiments to resolve the ambiguities. Therefore, the approach not only increases the demand for NMR instrument time but may elaborate the analysis as well. Of-course, the $^{15}$N, $^1$H$^N$ resolved higher dimensionality ($\geq$ 3D) NMR experiments (e.g. $^{15}$N/$^{13}$C edited 4D HNCOCANH) can easily resolve the problems arising because of overlapped resonances, however, a 4D/5D spectrum with reasonable high-resolution typically requires weeks-to-months time for data collection with conventional acquisition routines which severely limits the utility of NMR for studying unstable proteins. Though, the sampling problem can be easily circumvented using the non-uniform sampling approach -currently well established for processing 3D NMR data acquired with good signal-to-noise ratio [38]- but optimal processing of higher dimensionality ($\geq$ 3D) data has not yet been established and is not in routine as well [39]. Moreover, such type of data processing is computationally very demanding and aim at the final result through an iterative process that has to be performed with great care to minimize the artifacts [40].

However, reduced dimensionality (RD) NMR [41,42] offers the most viable and feasible solution to reduce the acquisition time by an order of magnitude. It provides higher dimensional information in a reduced dimensionality spectrum [43] through joint sampling of two or more chemical shifts in a single indirect dimension, thus provides considerable reductions in the measurement time. In addition to affording savings in measurement time, the sums and differences of chemical shifts -resulting from joint sampling- span a wider spectral range than the individual shifts themselves. However, the important consequence of the linear combination of chemical shifts is that it provides better dispersion and randomness to resolve the ambiguities arising because of degenerate chemical shifts while establishing the sequential connectivities in backbone assignment process [44,45]. The idea has been used earlier by Atreya et al [8] to solve the backbone assignment problem for proteins with very high shift degeneracy – based on 5D spectral information encoded in G$^2$FT NMR experiments. These experiments, although, make use of backbone $^{15}$N and $^{13}$C' shifts, that remain well dispersed for most of the folded and unfolded proteins, also include other nuclei with poorer dispersion, $^{13}$C$^{\alpha}$, $^{13}$C$^{\beta}$, $^{1}$H$^{\alpha}$, which may adversely affect the assignment process. As an alternative, a novel reduced dimensionality 3D NMR experiment –(5,3)D hNCO-CANH- has been presented here which makes full use backbone $^{15}$N and carbon ($^{13}$C' and $^{13}$C$^a$) chemical shifts. The resulted spectrum offers relatively higher chemical shift dispersion and randomness simultaneously (a) to break the backbone shift degeneracy and (b) to resolve the problems arising because of overlapping amide NH resonances. The performance of the experiment and the application of the method have been demonstrated here using (i) a 15.4 kDa size folded protein named tgADF (*Toxoplasma gondii* ADF, a 118 amino acid protein with an additional N-terminal twenty-one residue purification tag) and (ii) a 12 kDa size unfolded protein named CFP-10 (i.e. culture filtrate protein of *Mycobaterium tuberculosis*, a 100 amino acid protein with an additional C-terminal thirteen residue purification tag).

**Materials and Methods:**

The proposed reduced dimensionality experiment -(5,3)D-h<u>NCO</u>-<u>CAN</u>H was developed and tested successfully first using a standard $^{15}$N/$^{13}$C labeled protein sample of chicken SH3 domain (final concentration ~1.5 mM, dissolved in 50 mM Sodium phosphate buffer containing 90% $H_2O$/10% $D_2O$, pH 6.5 in high quality NMR tube sealed under inert atmosphere purchased from Cambridge Isotope Laboratories, Inc., USA: *http://www.isotope.com/cil/*). The application of the experiment has been demonstrated here on (i) a 15.4 kDa size folded protein referred here as tagged TgADF (i.e. *Toxoplasma gondii* ADF, a 118 amino acid protein with an additional N-terminal twenty-one residue purification tag from vector pET16b) expressed and purified as described here [46] without cleaving the additional twenty one residue purification tag and (ii) a 113 amino acid long unfolded protein referred here as tagged CFP-10 (i.e. culture filtrate protein of *Mycobacterium tuberculosis* with an additional C-terminal thirteen residue purification tag from vector pET28b) expressed and purified as described here [47] without cleaving the additional thirteen residue purification tag. The purpose of keeping terminal tags in both the cases was to induce the spectral complexity. The final $^{13}$C/$^{15}$N labeled samples of both the proteins (1.0 mM in concentration) were prepared in 50 mM Sodium phosphate buffer (pH 6.0) containing 150 mM NaCl and 90% $H_2O$/ 10% $D_2O$. All the experiments have been performed on a Bruker Avance III 800 MHz NMR spectrometer equipped with a Cryoprobe (acquisition parameters and measurement times for the different experiments are provided in **Table S1** of the Supporting Information). Along each indirect dimension, frequency selection has been achieved using standard States-TPPI method [48] where quadrature detection has been performed for $^{15}$N signal. Since all the spectra are acquired in a RD manner [41,42], the raw data does not require any pre-processing and are processed using routine method used for processing conventional 3D FT-NMR spectra. All the NMR data was processed using Topspin 2.1 (Bruker software: http://www.bruker.com/) and analyzed using CARA [49].

## Results and Discussion:

*Pulse sequence and Magnetization Transfer in (5,3)D-h(NCO)-(CAN)H experiment*

The pulse sequence for (5,3)D-h(NCO)-(CAN)H experiment has been shown in Electronic Supplementary Material (ESM, **Fig. S1**). It has been derived from the previously described HN(C)N [16] pulse sequence by tweaking both the indirect evolutions involving $^{15}$N nuclei (i.e. $t_1$ and $t_2$) according to reduced dimensionality (RD) NMR approach [41,42]. **Figure 1** traces the magnetization transfer pathway along with the respective frequency labeling schemes in this new pulse sequence. As explicitly depicted in **Fig. 1,** the $t_1$ evolution involves joint sampling of backbone $^{15}$N$_i$ and $^{13}$C'$_{i-1}$ chemical shifts, whereas $t_2$ evolution involves joint sampling of backbone $^{15}$N$_i$ and $^{13}$C$^\alpha_{i/i-1}$ chemical shifts. For each indirect dimension, $^{15}$N chemical shifts are detected in quadrature whereas backbone carbon ($^{13}$C' or $^{13}$C$^\alpha$) chemical shifts modulate the transfer amplitude. Depending upon the carbon ($^{13}$C' or $^{13}$C$^\alpha$) chemical shift involved in $^{15}$N frequency modulation, the $F_1$ and $F_2$ dimensions have been referred here as NC' and NC$^\alpha$. The theoretical description of the pulse sequence goes identically to that of HN(C)N [11,16], except that the frequencies along both the indirect dimensions will correspond to the sum and difference of frequencies of the jointly sampled nuclei. Though, the experiment encodes 5D spectral information, it leads to a three dimensional (3D) spectrum. Further, it preserves all the beneficial features of basic HN(C)N spectrum i.e. the presence of opposite signs for self and sequential correlation peaks and unidirectional sequential walk. Therefore, the resulted spectrum has been referred here as pseudo 5D HN(C)N.

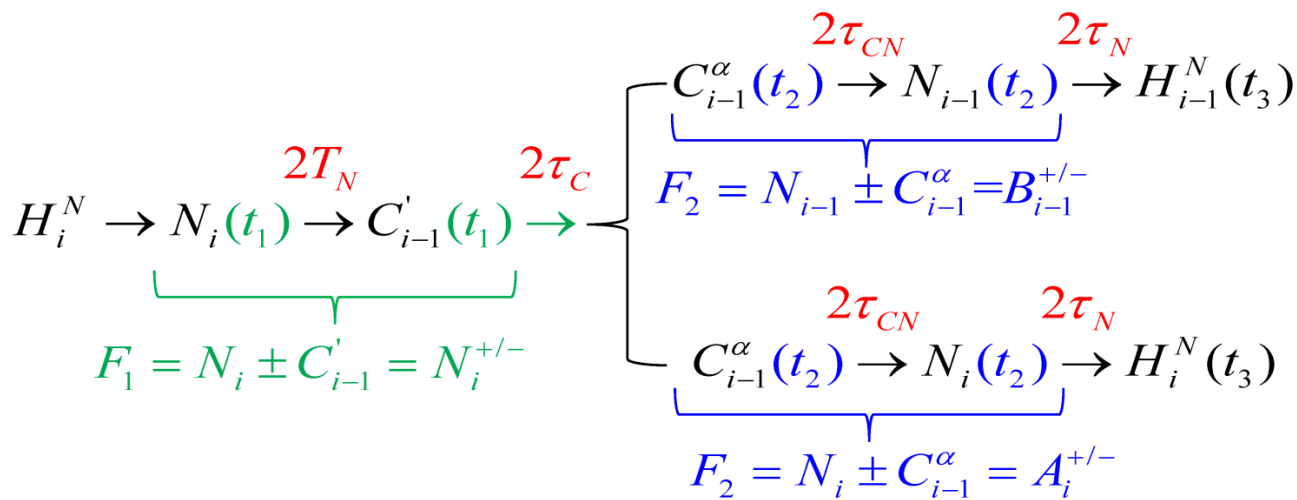

**Figure 1:** Schematic illustrations of the magnetization transfer pathway along with the frequency labeling of the appropriate nuclei employed in reduced dimensionality tailored HN(C)N experiment –(5,3)D- h(NCO)-(CAN)H. The delays $-2T_N, 2\tau_C, 2\tau_{CN},$ and $2\tau_N$ – during which the transfers indicated by the arrows take place in the pulse sequence– were set to 28, 9, 25 and 27 ms, respectively.

**Spectral Features and Assignment Strategy:**

Schematic three-dimensional (5,3)D- h(NCO)-(CAN)H spectrum and the correlations observed in the $F_1$(NC')-$F_3$($^1$H) and $F_2$(NCa)-$F_3$($^1$H) planes are shown in **Figure 2**. As depicted, the peaks appear at the following coordinates in the $F_1$-$F_3$ and $F_2$-$F_3$ planes of spectrum:

$$F_2 = A_i^+ / A_i^-, \ (F_1, F_3) = (H_i, N_i^+), (H_i, N_i^-) \tag{1}$$

$$F_2 = B_{i-1}^+ / B_{i-1}^-, \ (F_1, F_3) = (H_i, N_i^+), (H_i, N_i^-) \tag{2}$$

$$F_1 = N_i^+ / N_i^-, \ (F_2, F_3) = (H_i, A_i^+), (H_i, A_i^-), (H_{i-1}, B_{i-1}^+), (H_{i-1}, B_{i-1}^-) \tag{3}$$

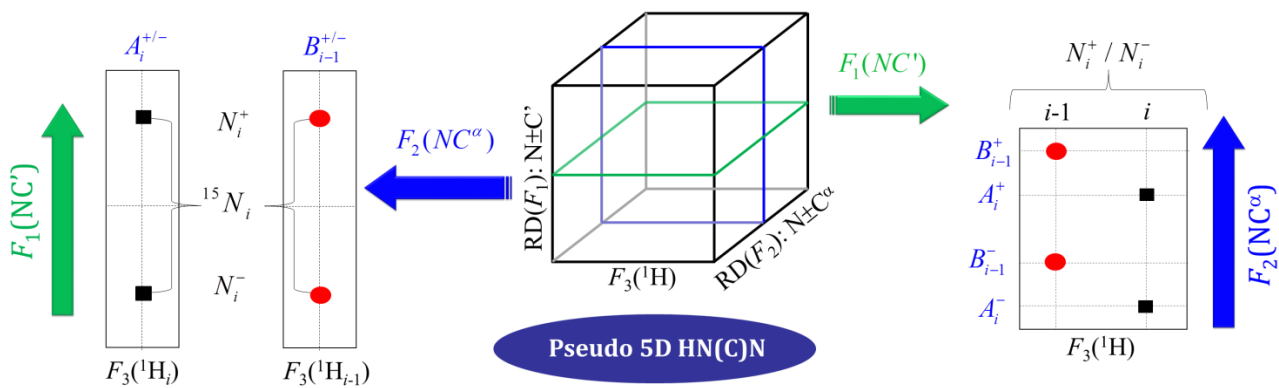

**Figure 2:** Schematic representation of the three-dimensional (5,3)D-h(NCO)-(CAN)H spectrum. The correlations observed in the $F_1$(NC')–$F_3$($^1$H) plane at the $F_2 = A_i^+ / A_i^-$ and $F_2 = B_{i-1}^+ / B_{i-1}^-$ (where $i$ correspond to a residue number) are shown on left side and the correlations observed in the $F_2$(NC$^\alpha$)–$F_3$($^1$H) planes at $F_1 = N_i^+ / N_i^-$ are shown on the right side. Squares and circles represent the self and sequential peaks, respectively. Black and red represent positive and negative phase of peaks, respectively.

The letter 'H' refers here to amide $^1$H$^N$ chemical shift and subscript '$i$' corresponds to the residue number. The notions used in **Eqns. 1-3** i.e. $N_i^+, N_i^-, A_i^+, A_i^-, B_{i-1}^+$ and $B_{i-1}^-$, refer to the linear combination of backbone $^{15}$N and $^{13}$C chemical shifts (the red color has been used to highlight the opposite phase of the peak in the final spectrum). Depending upon the offset of RF pulses used along $^{13}$C channel (i.e. $^{13}$C$^\alpha$_{offset} and $^{13}$C'_{offset}), the values for these notions evaluated as per the reduced dimensionality NMR convention [41,42] are:

$$N_i^+ = {}^{15}N_i^{obs} + \kappa*({}^{13}C'_{i-1} - {}^{13}C'_{offset}) \qquad (4)$$

$$N_i^- = {}^{15}N_i^{obs} - \kappa*({}^{13}C'_{i-1} - {}^{13}C'_{offset}) \qquad (5)$$

$$A_i^+ = {}^{15}N_i^{obs} + \kappa*({}^{13}C_{i-1}^\alpha - {}^{13}C_{offset}^\alpha) \qquad (6)$$

$$A_i^- = {}^{15}N_i^{obs} - \kappa*({}^{13}C_{i-1}^\alpha - {}^{13}C_{offset}^\alpha) \qquad (7)$$

$$B_{i-1}^+ = {}^{15}N_{i-1}^{obs} + \kappa*({}^{13}C_{i-1}^\alpha - {}^{13}C_{offset}^\alpha) \qquad (8)$$

$$B_{i-1}^- = {}^{15}N_{i-1}^{obs} - \kappa*({}^{13}C_{i-1}^\alpha - {}^{13}C_{offset}^\alpha) \qquad (9)$$

The symbol "$k$" here is the scaling factor and is set equal to $\gamma(^{13}C)/\gamma(^{15}N)=2.48$ in the present study. As shown in **Figure 2**, the $F_1$-$F_3$ plane corresponding to $F_2 = A_i^+ / A_i^-$ shows two correlation peaks **(Fig. 1C, left panel)**: one up-field $[H_i, N_i^+]$ and the other down-field $[H_i, N_i^-]$ from the $^{15}$N offset ($^{15}$N$_{offset}$), respectively, due to addition ($N_i^+$) and subtraction ($N_i^-$) of $^{15}$N$_i$ and $^{13}$C$_{i-1}$ chemical shifts (where $i$ is residue number). These correlation peaks have been referred here as self-peaks –notions used as per the HN(C)N convention. Similarly, the $F_1$-$F_3$ plane corresponding to $F_2 = B_{i-1}^+ / B_{i-1}^-$ shows two correlation peaks: $[H_{i-1}, N_i^+]$, and $[H_{i-1}, N_i^-]$, which have been referred here as sequential correlation peaks. On the other hand, the $F_2$-$F_3$ plane corresponding to $F_1= N_i^+ / N_i^-$, shows (a) two intra-residue correlation peaks ($B_{i-1}^+ / B_{i-1}^-$) referred here as sequential peaks and (ii) two inter-residue correlation peaks ($A_i^+ / A_i^-$) referred here as self-peaks **(Fig. 1C, right panel).** These notions are used as per the HN(C)N convention i.e. self-peaks are the ones which involve $^{15}$N$_i$ chemical shift, while sequential peaks involve $^{15}$N$_{i-1}$ chemical shift **(Eqns. 6-9)**. These self and sequential correlation peaks in different planes appear in opposite signs except when there is a Glycine at $i$ or $i$-1 position, thereby allowing identification of spin systems corresponding to glycines or following glycines in the sequence. These features are similar to HN(C)N spectrum, therefore, the new experiment has also been referred here as the pseudo 5D HN(C)N. Basically, it preserves all the beneficial features of basic HN(C)N spectrum [11,25] like e.g. (a) self and sequential correlation peaks have opposite signs (except in some special situations) and thus can be discriminated easily without involving any complementary experiment and (b) special patterns of positive and negative peaks appear around glycines and prolines and they serve as important start/check points during the course of sequential assignment walk; therefore explicit side chain assignments become less crucial for unambiguous backbone assignment. **Figure 3** shows the experimental demonstration of these features for the $F_1$(NC')-$F_3$($^1$H) and $F_2$(NCa)-$F_3$($^1$H) planes of pseudo 5D HN(C)N spectrum acquired on intrinsically unstructured protein CFP-10.

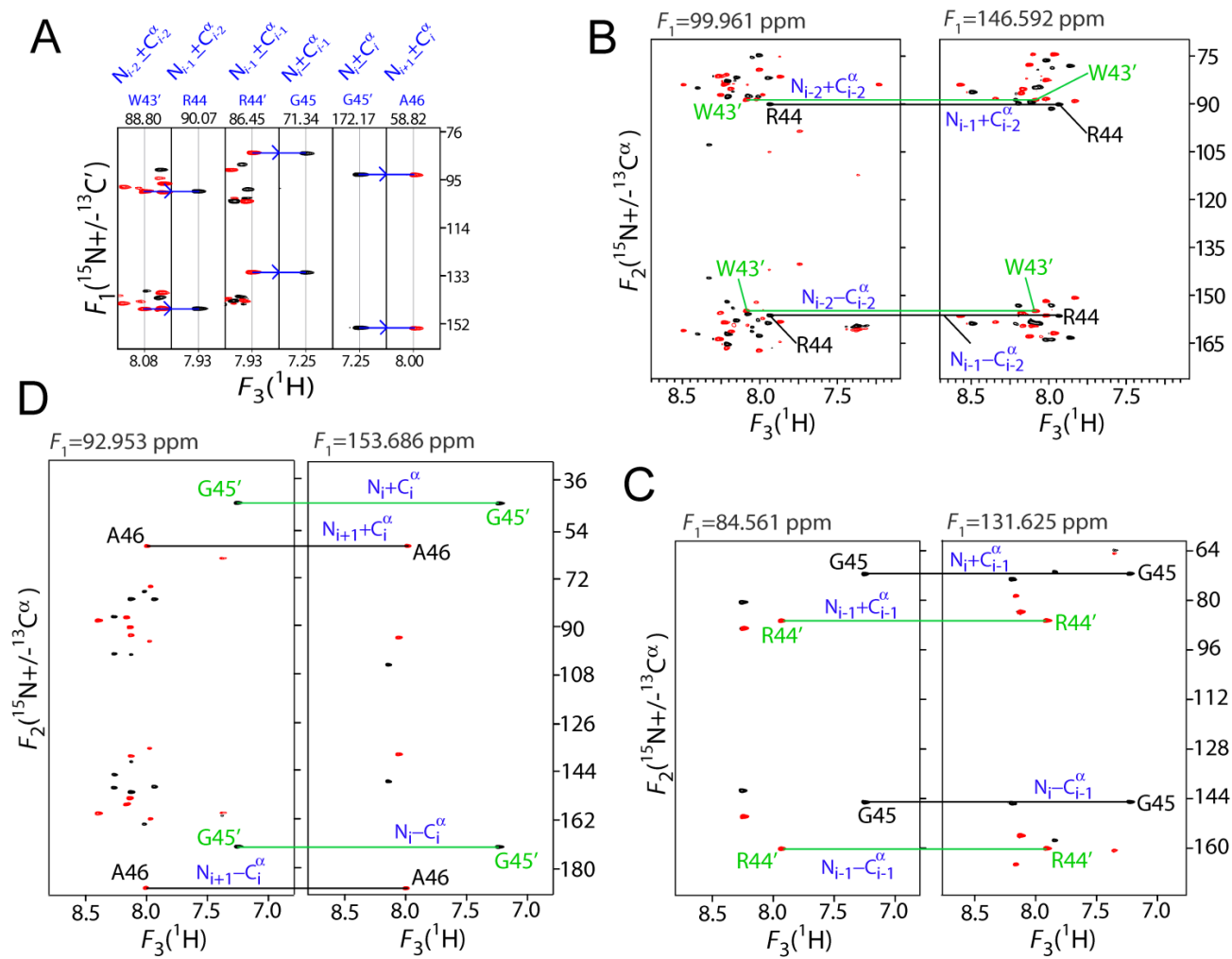

**Figure 3:** An illustrative example showing use of correlations observed in the $F_2$–$F_3$ planes of (5,3)D-h(<u>NCO</u>)-(<u>CAN</u>)H spectrum of unfolded CFP-10 protein at $N_i^{+/-} (= N_i +/- C'_{i-1})$ chemical shifts of residue $i$ for establishing a sequential ($i \rightarrow i$-1) connectivity between the backbone spin-systems [i.e. $A_i^+/A_i^- \rightarrow B_{i-1}^+/B_{i-1}^-$]. Red and black contours represent positive and negative phase of the peaks, respectively. As evident, a sequential spin-system ($B_{i-1}^+$ and $B_{i-1}^-$) –appearing in both the $F_2$–$F_3$ planes of the spectrum at $N_i^+ / N_i^-$ chemical shifts– can be easily identified because of its opposite peak sign compared to that of the self spin-system $A_i^+$ and $A_i^-$. This feature significantly reduces the search for sequential correlations.

Another important point need to be discussed here is the experiment time requirement. Compared to the basic HN(C)N experiment, the proposed RD experiment requires longer experimental time. This is due to the fact that the sums and differences of chemical shifts resulting from joint sampling span a wider spectral range than the individual shifts; therefore to achieve the same spectral resolution, one has to acquire more data points along the co-evolved dimensions with increased spectral width (generally two to three times of the primary $^{15}$N spectral width). However, long experiment time requirement may limit the utility of the method for studying proteins unstable in solution –either they start precipitating or degrading inside the NMR sample tube within matter of days. In such situations, the rapid data collection is very crucial and therefore, the experiment has also been modified according to L-optimization approach [50] for rapid data collection. The modification has been made similar to that implemented by Diercks *et. al.* [51] where aliphatic/water $^1$H magnetization is preserved maximally along +z-axis with minimal perturbation. This is accomplished by: (1) removing $^1$H decoupling with 180° inversion pulses and (2) placing band selective $^1$H radiofrequency pulses with excitation in the region of -1 to 5.5 ppm at the start of the pulse sequence and immediately after the second 90 ° pulse on $^1$H. The (5,3)D hNCO-CANH pulse sequence thus modified according to L-optimization approach has been shown in Supplementary material **(Fig. S1B)** and allows significant reduction in overall experiment time through speeding up the longitudinal relaxation of amide protons.

**Improved Peak Dispersion and Randomness to break the Amide Shift Degeneracy:**

As evident from the description of pulse sequence, the current experiment is basically a modification of previously described HN(C)N experiment [16]. In basic HN(C)N experiment, the use of amide $^{15}$N chemical shifts along both the indirect dimensions fails to resolve the NH overlap problems and therefore adversely affect the assignment when there is very high degree of backbone amide shift degeneracy. A solution to this problem has been presented here in the form of reduced dimensionality tailored HN(C)N experiment. Like basic HN(C)N, the experiment involves the use of amide $^{15}$N chemical shifts along both its indirect dimensions, however additionally, now these are modulated with the frequency of backbone carbon ($^{13}$C' and $^{13}$C$^a$) chemical shift to resolve the ambiguity involved in connecting the NH(i) and NH(i-1) resonances for overlapping NH peaks. The implementation is based on the fact that the linear combinations of backbone $^{15}$N and carbon ($^{13}$C$^\alpha$ and $^{13}$C') chemical shifts provide dispersion better compared to the dispersion of the individual chemical shifts. The fact has depicted in **Fig. S2** (see ESM) using average $^{15}$N, $^{13}$C$^\alpha$ and $^{13}$C' chemical shifts derived from the BMRB database [52] for all the 20 amino acids. As shown in the figure, the average $^{13}$C', $^{15}$N, and $^{13}$C$^\alpha$ chemical shifts (evaluated in terms of $^{15}$N frequency scale of 800 MHz NMR spectrometer, where 1 ppm = 80 Hz) individually show dispersions of ~20.0 (=8.0*2.48), ~21.5 and ~53.7 (21.5 x 2.48) ppm, respectively, **(Fig. S2A, S2B and S2C)**. However, the linear combination of these chemical shifts provides dispersion relatively higher than the individual chemical shifts: (i) the subtraction and addition of $^{15}$N$_i$ and $^{13}$C'$_j$ chemical shifts (*i* and *j* represent the amino acid type) provide dispersion of 35.4 ppm and 35.5 ppm, respectively, **(Fig. S2D)**, (ii) the subtraction and addition of $^{15}$N$_i$ and $^{13}$C$^\alpha_j$ chemical shifts provide dispersion of 69.5 ppm and 68.0 ppm, respectively **(Fig. S2E)** and (iii) the subtraction and addition of

$^{15}$N$_i$ and $^{13}$C$^{\alpha}_i$ chemical shifts provide dispersion of 67.7 ppm and 59.4 ppm, respectively **(Fig. S2F).** Experimentally, we have also demonstrated this fact on folded tagged TgADF protein (see Supplementary Material, **Fig. S3**). Besides higher shift dispersion, the additional advantage here is that these dispersions of peaks in the sum and difference frequency regions are different. Therefore, the simultaneous analysis of both the spectral halves greatly helps to resolve the ambiguities e.g. if there in an ambiguity because of degeneracy in additional half of the spectrum, subtraction half of the spectrum helps to resolve it and vice-versa.

As shown in **Figure 3**, the sum and difference frequencies are well separated in their respective co-evolved dimensions. However, to achieve separation between these two sets of frequencies (i.e. $N_i^+/A_i^+/B_{i-1}^+$ and $N_i^-/A_i^-/B_{i-1}^-$) should be careful about (i) the spectral width used along the co-evolved dimensions (NC' and NCa) as the linear combinations of chemical shifts span increased spectral range and (ii) the offset frequencies used along $^{13}$C' and $^{13}$C$^a$ channels. To evaluate these parameters, a small exercise was performed on average $^{15}$N and $^{13}$C$^{\alpha}$/$^{13}$C' chemical shifts of all the common 20 amino acids (data taken from BMRB) using **Eqns 4-9**. The analysis revealed that for a given protein the two frequency regions (i.e. corresponding to sum and difference of the chemical shifts) will be separated from each other (i) if the $^{13}$C carrier is kept 1–2 ppm away from the most down fielded shifted $^{13}$C chemical shift; this is typically close to: ~184-188 ppm for $^{13}$C' channel during $t_1$ evolution and ~68-70 ppm for $^{13}$C$^{\alpha}$ channel during $t_2$ evolution and (ii) the spectral widths are increased along the co-evolved $F_1$(NC') and $F_2$(NCa) dimensions have to be increased accordingly; typically close to ~90 and ~160 ppm, respectively. Accordingly, the final spectra were recorded using acquisition parameters as shown in **Table S1** of Supplementary Material**.** Another point to be discussed here is that the proposed double reduced dimensionality experiment requires long measurement time compared to conventional 3D NMR experiments which may limit its utility for studying proteins unstable in solution. However as the sequence does not involve water and aliphatic protons in its coherence transfer pathway, the long experiment time problem has been circumvented easily by exploiting the H-flip approach to achieve longitudinal relaxation enhancement [50] for rapid data collection; thus rendering the method practical for routine NMR studies of small (MW < 15 kDa) folded and a range of intrinsically unfolded proteins. A demonstration of the L-optimized pseudo 5D HN(C)N experiment on a ~15.4 kDa size folded protein referred here as tagged TgADF has been shown in Supplementary material **(Fig. S3)**.

*Assignment Strategy*

**Figure 4** depicts the stepwise protocol for unambiguously establishing the sequential (*i* to *i*-1) connectivities between backbone spin systems using the current (5,3)D HN(C)N experiment in concert with the established HNCO [53] and other recently proposed reduced dimensionality experiments: (3,2)D-H<u>NCO</u> [54], (3,2)D-H<u>N</u>co<u>CA</u> [54] and (4,3)D CBCA<u>CO</u>NH [55]. Briefly, the procedure involves: **(a)** identification of all the backbone [H$_i$, N$_i$, C'$_{i-1}$] spin systems (where *i* is residue number) using the HSQC spectrum in concert with the established HNCO spectrum, **(b)** the information is used to label the peaks in the (3,2)D H<u>NCO</u> spectrum: one up-field [$H_i, N_i^+$] and the other down-field [$H_i, N_i^-$] from the $^{15}$N offset ($^{15}$N$_{offset}$), respectively, resulting due to addition ($N_i^+$) and subtraction ($N_i^-$) of $^{15}$N$_i$ and $^{13}$C'$_{i-1}$ chemical shifts (see **Eqns 4 and 5**), **(c)** next using $^{15}$N+/-$^{13}$C' edited experiment -(4,3)D CBCACONH- the sequential (*i*-1) $^{13}$C$^\alpha$ and $^{13}$C$^\beta$ chemical shifts are identified for each backbone [H$_i$, N$_i$, C'$_{i-1}$] spin system, **(d)** the information is then used to label the peaks in the (3,2)D-H<u>N</u>co<u>CA</u> spectrum: one up-field [$H_i, A_i^+$] and the other down-field [$H_i, A_i^-$] from the $^{15}$N offset ($^{15}$N$_{offset}$), respectively, resulting due to addition ($A_i^+$) and subtraction ($A_i^-$) of $^{15}$N$_i$ and $^{13}$C$^\alpha_{i-1}$ chemical shifts (see **Eqns 6 and 7**), **(e)** now the $F_3$($^1$H)-$F_2$(NCa) planes of the (5,3)D HN(C)N spectrum corresponding to $F_1(NC') = N_i^+ / N_i^-$ are visually inspected for identifying the sequential $B_{i-1}^+ / B_{i-1}^-$ correlations. These correlations are identified based on two criteria: first by their opposite signs compared to the self-correlation ($A_i^+ / A_i^-$) peaks (except when there is glycine at *i* or *i*-1 position, in such situations the self and sequential correlations will be of same sign) and second these will be present on both the $F_3$($^1$H)-$F_2$(NCa) planes of spectrum corresponding to sum ($N_i^+$) and difference ($N_i^-$) of $^{15}$N$_i$ and $^{13}$C'$_{i-1}$ chemical shifts, **(f)** each sequential connectivity identified in step '**e**' is further validated by matching the modulated $^{15}$N chemical shifts (i.e. $N_i^+$ or $N_i^-$) on $F_3$($^1$H)-$F_2$(NCa) planes of spectrum corresponding to self ($A_i^+ / A_i^-$) and sequential ($B_{i-1}^+ / B_{i-1}^-$) chemical shifts, **(g)** once all the possible sequentially connections are established between the various backbone spin systems, the stretches of sequential connectivities are transformed into the final assignment. For this purpose, the amino acid type identification based on $^{13}$C$^\beta$ chemical shifts is carried out. These chemical shifts provide unique identification of spin systems corresponding to glycines, alanines and serines/threonines which in turn serve as important check points for mapping the stretches of sequentially connected HSQC peaks on to the primary sequence for assigning them sequence specifically.

Though, the proposed sequential assignment method is slightly tricky, helps to reduce the probability of an ambiguous assignment in situations when there is high degree of amide shift degeneracy. As evident from the assignment protocol depicted in **Figure 4**, the pseudo 5D HN(C)N spectrum provides an alternative route for establishing the sequential (*i* to *i*-1) connectivity between the backbone spin systems. The strategy does not involve the use of backbone $^{15}$N, $^{13}$C$^a$, and $^{13}$C' shifts directly, rather modulated backbone $^{15}$N chemical shifts i.e. $N_i^{+/-}/A_i^{+/-}$ or $N_j^{+/-}/A_j^{+/-}$ **(Eqns 4-7)** are used to confirm the sequential (*i* to *i*-1) connectivity between two spin systems. As the modulation depends upon the backbone carbon ($^{13}$C$^\alpha$ or $^{13}$C') chemical shifts, therefore if the overlapped spin systems (i.e. when $H_iN_i=H_jN_j$) differ even slightly in their sequential carbon chemical shifts, can be connected to their true neighboring partners unambiguously. Therefore, we believe that the approach would greatly facilitate the sequential assignment walk through resolving the ambiguities arising because of overlapped amide NH cross peaks. Overall, the double RD version of HN(C)N experiment mimics the G$^2$FT concept [8] to resolve the amide shift degeneracy requires. However, there is a slight difference in the assignment strategy. The one based on G$^2$FT type of experiments [8] requires analysis of pair-wise complementary spectra from two such experiments for establishing the sequential connectivities, whereas the pseudo 5D HN(C)N spectrum does not involve any other complementary experiment of (5,3)D type. Rather, (4,3)D CBCA*CON*H spectrum is used simultaneously (i) to obtain the unambiguous amino acid type identification for all the possible [$H_i,N_i,C_{i-1}$] spin systems and (ii) to differentiate the self and sequential correlations in pseudo 5D HN(C)N spectrum making use of the linear combinations of backbone $^{15}$N(*i*) and $^{13}$Ca(*i*-1) chemical shifts.

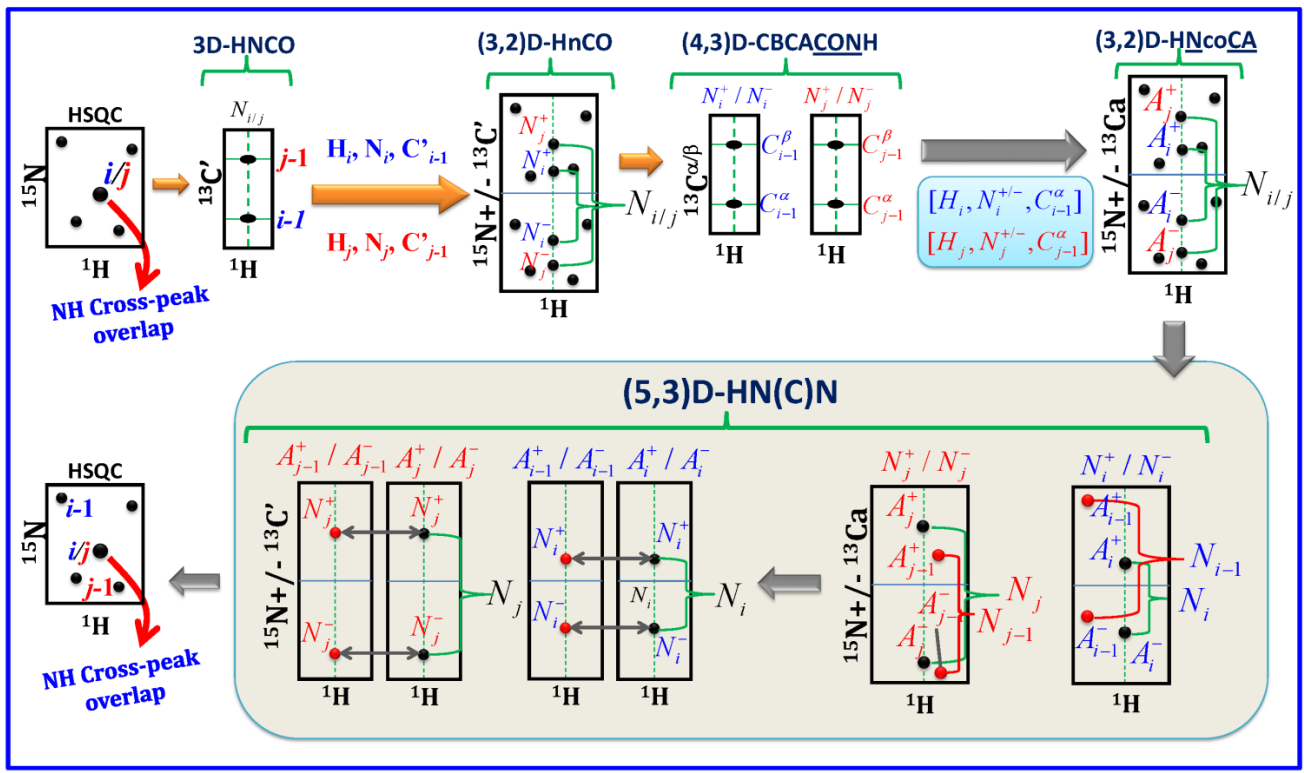

**Figure 4:** Schematic illustrating the strategy of establishing the sequential (*i* to *i*-1) connectivity based on the reduced dimensionality (5,3)D HN(C)N experiment in combination with standard HNCO experiment and other established reduced dimensionality triple resonance NMR experiments like (3,2)D-H<u>N</u>CO, (4,3)D-CBCA<u>CO</u>NH, and (3,2)D-H<u>N</u>co<u>CA</u>. Red and black colors in (5,3)D HN(C)N represent positive and negative phase of peaks, respectively. Blue line in (3,2)D-H<u>N</u>CO, (3,2)D-H<u>N</u>co<u>CA</u> and (5,3)D HN(C)N spectral planes represent the $^{15}$N offset. As evident, the strategy based upon the use of modulated backbone chemical shifts i.e. $N_i^{+/-}/A_i^{+/-}$ or $N_j^{+/-}/A_j^{+/-}$ **(Eqns 4-9)** provides unambiguous identification of sequential NH cross peak coordinates i.e. (H$_{i-1}$N$_{i-1}$) and (H$_{j-1}$N$_{j-1}$) for the overlapped amide cross peaks of the HSQC spectrum.

### *Application to Intrinsically Disordered Protein CFP-10:*

Assignment of intrinsically disordered proteins (IDPs) is seriously hampered by (a) the extensive amide cross peak overlap and (b) the poor backbone carbon chemical shift dispersion especially of $^{13}C^{\alpha}$ and $^{13}C^{\beta}$ spins. The recently proposed reduced dimensionality tailored (4,3)D HN(C)N experiments [44,45] can easily circumvents the problems arising because of poor dispersion and degeneracy of carbon chemical shifts. However, in situations when the HSQC spectrum has severe spectral crowding and cross-peak overlap, the $^{15}$N chemical shift based assignment process becomes very difficult and problematic. To address this issue, we propose here to use the pseudo 5D HN(C)N experiment in combination with HNCO and other established reduced dimensionality NMR experiments like (3,2)D-H<u>N</u>CO, (4,3)D-CBCA<u>CO</u>NH, and (3,2)D-H<u>N</u>co<u>CA</u>. We used these experiments for the backbone assignment of a 113 amino acid unfolded tagged CFP-10 protein **(Fig 5A)** using the protocol shown in **Fig. 4**. The finger-print $^1$H-$^{15}$N HSQC spectrum of the protein is shown in **Figure 5B**. The limited amide $^1$H chemical shift dispersion (ranging from 7.7 to 8.7 ppm) indicated that the protein is largely unstructured in solution. As the sequence does not contain any proline **(Fig 5A),** a total 113 cross peaks from backbone amide NH pairs were expected in the $^1$H-$^{15}$N HSQC spectrum. However, the peak picking analysis discerned only 92 isolated peaks above the noise level indicating significant amide cross peak overlap in the HSQC spectrum. One such situation for the two glycine residues (Gly37 and Gly80) has been shown in **Figure 5C**. However, the HNCO spectrum clearly shows that there are two peaks overlapping in that spectral region of the $^{15}$N HSQC spectrum. As these overlapped peaks differ slightly in their sequential (*i*-1) carbonyl carbon ($^{13}$C') chemical shifts **(Fig 5D)**, therefore these can be resolved spatially in the (3,2)D H<u>N</u>CO spectrum (see **Fig 5E** and **5F,** representing the addition and subtraction halves of the spectrum, respectively). Based on this fact, the combined analysis of HSQC and 3D HNCO spectra were performed which led to the identification of 109 spin systems (nearly close to the expectation value of 113). Of them, 21 spin systems were found to be in overlapped condition. Next, the ($^{15}$N+/-$^{13}$C') edited (4,3)D-CBCA<u>CO</u>NH experiment [55] was used for unambiguous identification of sequential (*i*-1) backbone $^{13}C^{\alpha}$ and $^{13}C^{\beta}$ chemical shifts for all the backbone spin systems **(Fig 5H)**. As evident from **Figure 5G**, the conventional $^{15}$N resolved CBCAcoNH experiment fails to provide such unambiguous information for severely overlapped spin systems. Finally, the linear combinations of backbone amide and (sequential) carbon chemical shifts obtained for various spin systems are used to establish the sequential connectivities according to the protocol illustrated in **Figure 5**. An example demonstration of the sequential assignment walk based on pseudo 5D HN(C)N spectrum is shown in **Figure 3**. The demonstration is shown for the series of residues Trp43-Arg44-Gly45 of CFP-10. Following this novel integrated approach, we were able to accomplish a nearly complete sequence specific backbone assignment of tagged CFP-10 protein except for first two N-terminal residues (which were not observed in the HSQC spectrum). The obtained backbone ($^{1}$H, $^{15}$N, $^{13}C^{\alpha}$, $^{13}C^{\beta}$ and $^{13}$C') chemical shift assignments have been deposited in the BioMagResBank (BMRB; http://www.bmrb.wisc.edu) under the BMRB accession number **19507**.

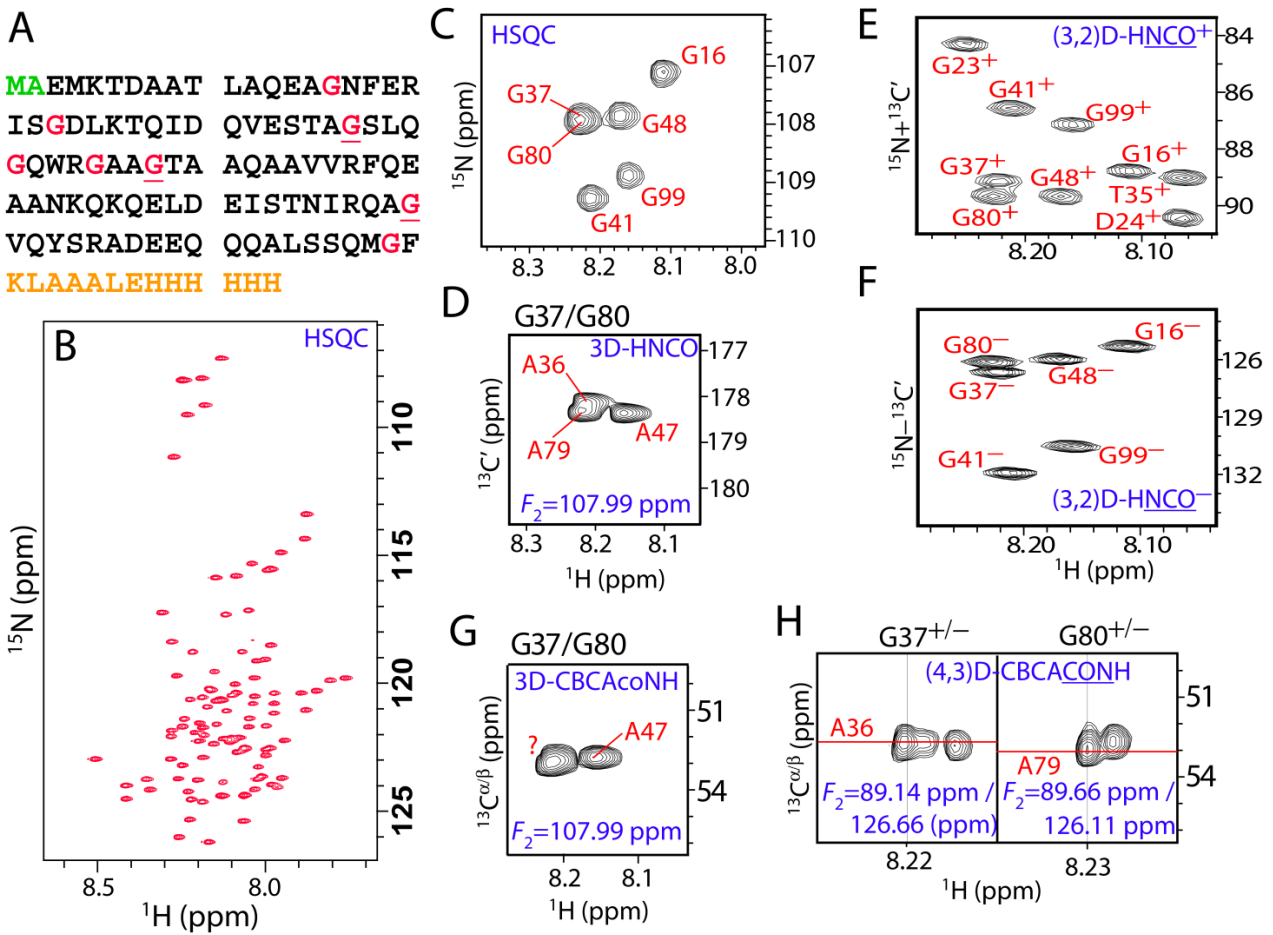

**Figure 5: (A)** Amino acid sequence of the tagged CFP-10 protein used in this study where amino acids colored in yellow represent the C-terminal purification tag. **(B)** $^{1}$H-$^{15}$N HSQC spectrum of the protein at 290 K and pH 6.5. **(C)** A part of $^1$H-$^{15}$N HSQC spectrum showing NH cross peak overlap (for residues Gly37 and Gly80) and amide $^{15}$N shift degeneracy for three glycine residues (i.e. Gly37, Gly48, and Gly80); all following alanines in the sequence (i.e. Ala36, Ala47, and Ala79). Such situations lead to ambiguities while establishing the sequential connectivities between the backbone amide (NH) cross peaks using experiments which involve backbone amide resonances. However, the proposed strategy simply resolves such ambiguities if the sequential (*i*-1) backbone carbon chemical shifts (i.e. $^{13}$C' and $^{13}$C$^\alpha$) for such overlapped HSQC peaks are even slightly different. **(D)** $F_1(^{13}C')$-$F_3(^1H)$ plane of the standard HNCO spectrum at $F_2(^{15}N)$=107.99 ppm (corresponding to residues Gly37 and Gly80 overlapping in the HSQC spectrum). **(E** and **F)** spectral regions from the addition and subtraction halves of the (3,2)D-H<u>NCO</u> spectrum showing the separation of modulated amide cross-peaks corresponding to residues Gly37 and Gly80. **(G)** $F_1(^{13}C^{a/b})$-$F_3(^1H)$ plane of the conventional CBCAcoNH spectrum at $F_2(^{15}N)$=107.99 ppm (corresponding to residues Gly37 and Gly80 overlapping in the HSQC spectrum). **(H)** $F_1(^{13}C^{a/b})$-$F_3(^1H)$ planes of the ($^{15}$N+/-$^{13}$C') edited (4,3)D-CBCA<u>CON</u>H experiment spectrum: left at $F_2(NC')$=89.14/126.66 ppm (corresponding to N$^{+/-}$ shifts of residues Gly37) and right at $F_2(NC')$=89.66/126.11 ppm (corresponding to N$^{+/-}$ shifts of residues Gly80).

**Concluding Remarks:**

In conclusion, a novel reduced dimensionality tailored HN(C)N experiment -(5,3)D h(<u>NCO</u>)(<u>CAN</u>)H– is presented to facilitate the sequence specific backbone resonance assignment of proteins possessing highly crowded and overlapped HSQC spectra e.g. unfolded proteins, alpha-helical proteins or folded proteins containing several disordered fragments. To resolve the problems arising because of amide and carbon chemical shift degeneracy, the experiment exploits the established co-evolution concept [41,42] to modulate these shifts in terms of their linear combinations. Overall, the experiment is superior both in terms of (a) peak dispersion resulting from addition and subtraction of chemical shifts and (b) efficacy with which a sequential connectivity between two backbone spin systems is established (or validated) unambiguously. Therefore, the experiment –like any other higher dimensionality 4D/5D triple resonance NMR experiment used in protein assignment- can be used for efficient and rapid sequential assignment in unfolded proteins and folded proteins with crowded HSQC spectrum.

The only limitation associated with the experiment is its low sensitivity as the experiment employs many steps of magnetization transfer via several long transfer delays and additional chemical shift evolution periods compared to the basic HN(C)N experiment. Therefore, the crucial requirement for the experiment to work is the slow relaxing (i.e. long $T_2$) backbone resonances. As the intrinsically disordered proteins often possess slow-relaxing NMR signals due to their inherent flexibility, the experiment would be of routine use [56]. For the same reason, the experiment can also be used to resolve the problems arising because of amide shift degeneracy in small-to-medium size folded proteins (MW ranging from 10-15 kDa) containing disordered fragments. In such proteins, mainly the disordered fragments are involved in generating the amide cross peak overlaps rendering the backbone assignment difficult and problematic. However because of inherent flexibility, these disordered fragments offers adequate sensitivity for the pseudo 5D HN(C)N spectrum to produce signals from these regions to resolve the ambiguities.

**Acknowledgement:**

This work is being financially supported by the Department of Science and Technology under SERC Fast Track Scheme (Registration Number: **SR/FT/LS-114/2011**) for carrying out the research work. DK would also like to acknowledge the Department of Science and Technology (DST), India for providing funds for the 800 MHz NMR spectrometer at Centre of Biomedical Research (CBMR), Lucknow, India.

# Supplementary Material:

## Pseudo 5D HN(C)N Experiment to Facilitate the Assignment of Backbone Resonances in Proteins Exhibiting High Backbone Shift Degeneracy

Dinesh Kumar,[1,#] Nisha Raikwal,[1] Vaibhav Kumar Shukla,[2] Himanshu Pandey,[2] Ashish Arora[2] and Anupam Guleria[1]

[1]*Centre of Biomedical Research (CBMR), SGPGIMS Campus, Raibareli Road, Lucknow-226014, India*
[2]*Molecular and Structural Biology Division, CSIR, Central Drug Research Institute, Lucknow 226 001, India*

[#]**Author for Correspondence:**

Dr. Dinesh Kumar
(Assistant Professor)
Centre of Biomedical Research (CBMR),
SGPGIMS Campus,
RB Road, Lucknow-226014
Uttar Pradesh, India
Mobile: +91-9044951791, +91-8953261506
Fax: +91-522-2668215
Email: dineshcbmr@gmail.com
Webpage: www.cbmr.res.in/dinesh.html

**KEYWORDS:** Reduced Dimensionality NMR; Backbone Resonance Assignment; HN(C)N; Pseudo 5D HN(C)N; Sequential Correlations.

**ABBREVIATIONS:** NMR, Nuclear Magnetic Resonance; HSQC, Heteronuclear Single Quantum Correlation; CARA, Computer Aided Resonance Assignment; RD: Reduced dimensionality; BMRB: Biological Magnetic Resonance Bank; ESM, Electronic Supplementary Material

**Figure S1:** RF Pulse sequences of a non-L **(A)** and L-optimized **(B)** (5,3)D-h<u>NCO</u>-<u>CA</u>NH experiment. Hollow and black solid rectangular bars in $^1$H/$^{15}$N channel represent, respectively, 90° and 180° hard (non-selective) pulses. Likewise, empty and solid lobes in carbon ($^{13}$C' or $^{13}$C$^\alpha$) channels indicate, respectively, 90° and 180° selective pulses, respectively. Unless indicated, the pulses are applied with phase x. Standard Gaussian cascade pulses [1] were used for band selective excitation and inversion along the $^{13}$C channel –shape Q3 for 180° inversion/refocusing (filled lobes) and shape Q5 for 90° excitation (hollow lobes). Proton decoupling using the DIPSI-2 decoupling sequence [2] with field strength of 6.3 kHz is applied during most of the $t_1$ ($^{15}$N/$^{13}$C') and $t_2$ ($^{15}$N/$^{13}$Ca) evolution periods, and $^{15}$N decoupling using the GARP-1 sequence [3] with the field strength 0.9 kHz is applied during acquisition. The carrier frequencies for pulses in $^{13}$C$^\alpha$ and $^{13}$C' channel were set at 54.0 ppm and 174 ppm, respectively, whereas the red vertical arrows labeled as 186 and 70 ppm indicate the offsets of carbon channel employed during $t_1$ evolution of $^{13}$C' nuclei and $t_2$ evolution of $^{13}$Ca nuclei, respectively. The encircled red pulse on $^{13}$C$^\alpha$ channel is crucial for tuning the experiment for generation of different check points in the final spectrum (like glycines/alanines and serines/threonines, for detail see these references [4,5]). Yellow pulses were applied for compensation of off-resonance effects (Bloch-Siegert phase shift) [6]. The values for the individual periods containing $t_1$ evolution of $^{15}$N nuclei are: $T_N^a = T_N/4 + t_1/4$ and $T_N^b = T_N/4 - t_1/4$. At the same time, the $^{13}$C' nuclei are co-evolved in a semi-constant time manner leading to the linear combination: $\Omega(^{15}N) \pm \kappa^*\Omega(^{13}C')$, where tunable scaling factor 'κ' was set to 2.48. The values for the individual periods containing $t_1$ evolution of $^{13}$C' nuclei are: $\tau_C^a = t_1/2$, $\tau_C^b = \tau_C$ and $\tau_C^c = \tau_C - t_1/2$. During $t_2$ evolution, the $^{15}$N nuclei are co-evolved with $^{13}$C$^\alpha$ nuclei in constant time manner. The values for individual periods containing $t_2$ evolution of $^{15}$N nuclei are: $\tau_N^a = \tau_N/4 - t_1/4$ and $\tau_N^b = \tau_N/4 + t_1/4$; and that of $^{13}$C$^\alpha$ nuclei are: A = $t_2/2$, B = $\tau_{CN} - \tau_C$, and C = $\tau_{CN} - t_2/2$. The other delays are set to λ = 2.5 ms, ε = 5.4 ms, δ = 2.5 ms, $\tau_C$ = 4.5 ms, $T_N$ = 14.0 ms, $\tau_{CN}$ = 12.5 ms and $\tau_N$ = 13.5 ms. The $\tau_{CN}$ must be optimized and is around 12-15 ms. The phase cycling for the experiment is $\Phi_1$ = 2($x$), 2(-$x$); $\Phi_2$ = $\Phi_3$= $x$, -$x$; $\Phi_4$ = $\Phi_5$ = $x$; $\Phi_6$ = 4($x$), 4(-$x$); and $\Phi_{receiver}$ = 2($x$), 4(-$x$), 2($x$). The frequency discrimination in $t_1$ and $t_2$ has been achieved using States-TPPI phase cycling [7] of $\Phi_1$ and $\Phi_5$, respectively, along with the receiver phase. States-TPPI based quadrature detection has been performed for $^{15}$N signals. The gradient (sine-bell shaped; 1 ms) levels are as follows: $G_1$=30%, $G_2$=30%, $G_3$=30%, $G_4$=30%, $G_5$=50%, $G_6$=80% and $G_7$=20% of the maximum strength 53 G/cm in the z-direction. The recovery time after each gradient pulse was 160 μs. Before detection, WATERGATE scheme [8] has been employed for better water suppression. Water saturation is minimized by keeping water magnetization

along the z-axis during acquisition with the use of water selective 90º pulses [9]. For L-optimized sequence **(B)** DIPSI-2 decoupling in $^1$H channel was replaced by two broadband inversion (BIP-720-50-20) pulses of 150 μs (colored in blue) and a single 180° hard pulse (colored in black). In L-optimized pulse sequence **(B)**, band selective EBURP-2 and a time-reversed band-selective EBURP-2 [10] each of 1.0 ms duration centered at 2.2 ppm (covering a range of -1 to 5.5 ppm) was applied on $^1$H channel before first 90° pulse and immediately after the second 90° pulse on $^1$H, respectively.

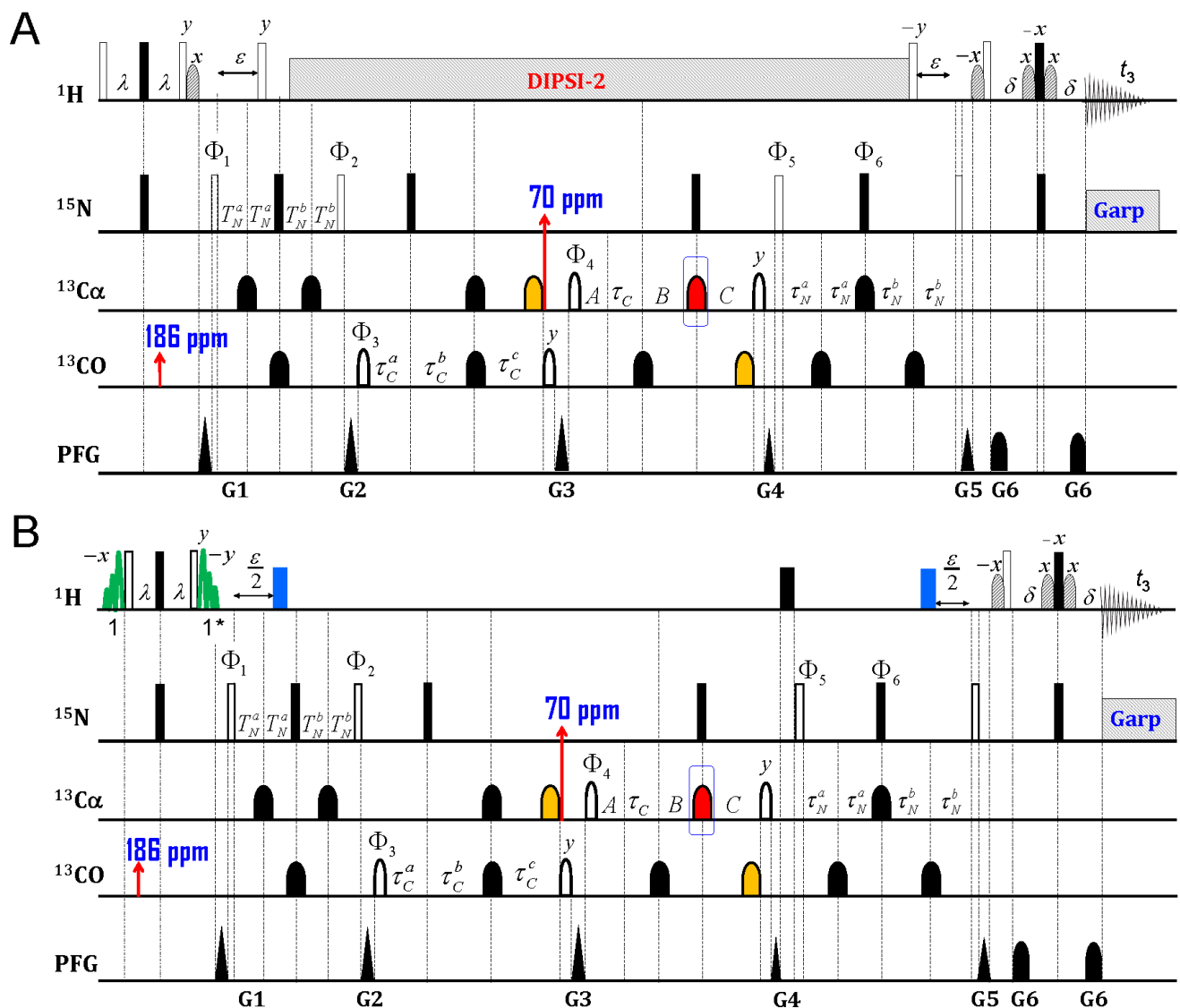

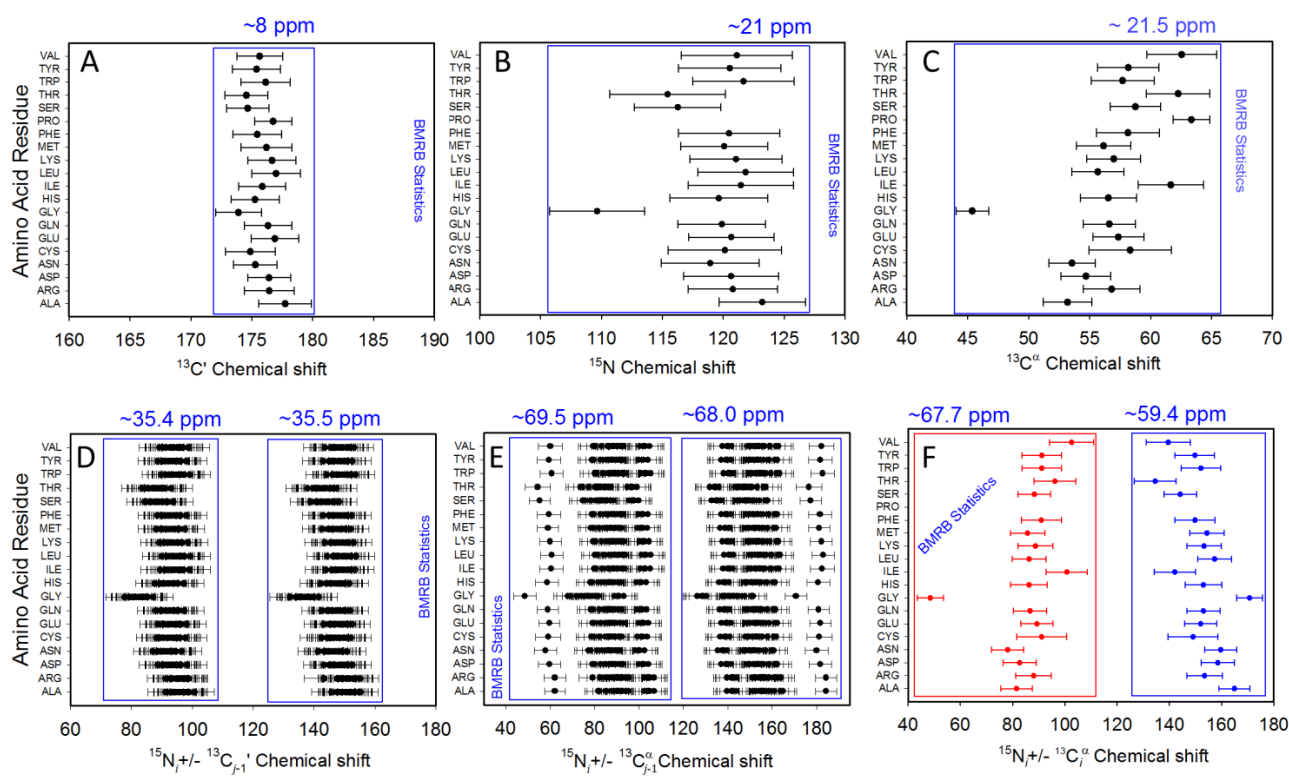

**Figure S2:** Comparison between the dispersions of individual $^{15}N/^{13}C^{\alpha}/^{13}C'$ chemical shifts and that for the linear combination of these chemical shifts for 20 common amino acids. In (A), (B) and (C) the average chemical shifts of $^{13}C'$, $^{15}N$, and $^{13}C^{\alpha}$ are plotted against the amino acid residue type. The average values and the respective standard deviations of backbone $^{15}N/^{13}C^{\alpha}/^{13}C'$ chemical shifts have been taken from the BMRB statistical table containing values calculated from the full BMRB database (http://www.bmrb.wisc.edu/ref_info/statful_.htm#1) [52]. This includes paramagnetic proteins, proteins with aromatic prosthetic groups, and entries where chemical shifts are reported relative to uncommon chemical shift references. Standard deviations in these values are plotted as error bars. In **(D)**, the linear combinations of $^{15}N_i$ and $^{13}C'_{j-1}$ average chemical shifts for all the amino acid types are shown evaluated according to **Eqns 4** and **5** (where *i* and *j* are any two amino acids). In **(E)**, the linear combinations of $^{15}N_i$ and $^{13}C^{\alpha}_{j-1}$ average chemical shifts for all the amino acid types are shown evaluated according to **Eqns 6** and **7**, (where *i* and *j* are any two amino acids). In **(F)**, the linear combinations of $^{15}N_i$ and $^{13}C^{\alpha}_i$ average chemical shifts for all the amino acid types are shown evaluated according to **Eqns 8** and **9**, (where *i* is any amino acid). The shifts shown in **1D, 1E** and **1F** have been calculated using $^{13}C'$ and $^{13}Ca$ offsets of 186.8 and 70 ppm, respectively. The deviations shown in **1D, 1E** and **1F** were calculated as: $\Delta\delta(NC)= [\Delta\delta(^{15}N)^2 – (\Delta\delta(^{13}C) \times 2.48)^2]^{1/2}$ where $\Delta\delta(^{15}N)$ and $\Delta\delta(^{13}C)$ are the standard deviations for the individual chemical shifts (plotted in **1A, 1B** and **1C**). The overall chemical shift dispersion achievable in each case has been shown at the top of each plot. Comparison clearly shows that the linear combinations of chemical shifts lead to better dispersions compared to the individual chemical shifts.

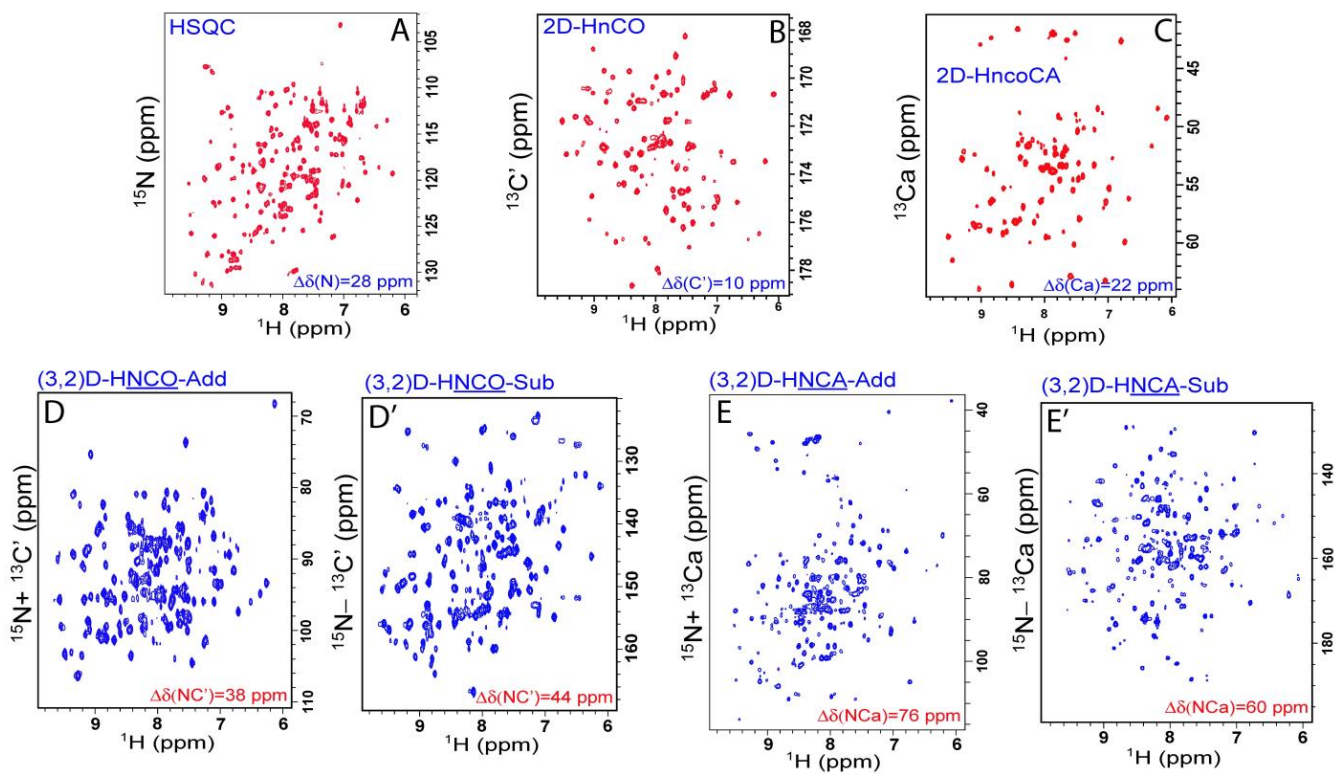

**Figure S3:** The overall dispersions of individual $^{15}N/^{13}C^{\alpha}/^{13}C'$ chemical shifts shown in comparison with that of their linear combinations obtained experimentally in case of $^{13}C/^{15}N$ labeled tgADF. **(A)** is the basic $^{1}H$-$^{15}N$ HSQC spectrum showing dispersion for the backbone amide $^{15}N$ shifts, **(B)** is the 2D-HnCO spectrum showing backbone carbonyl ($^{13}C'$) shift dispersion, **(C)** is the 2D-HncoCA spectrum showing backbone alpha-carbon ($^{13}C^{\alpha}$) shift dispersion, **(D)** and **(D')** represent the addition and subtraction halves of the (3,2)D-HNCO spectrum and show the dispersions available for the linear combination of backbone $^{15}N$ and $^{13}C'$ chemical shifts and **(E)** and **(E')** represent the addition and subtraction halves of the (3,2)D-HNCA spectrum and show the dispersions available for the linear combination of backbone $^{15}N$ and $^{13}C^a$ chemical shifts. Note that, the reduced dimensionality tailored $F_1$ dimensions in **(D), (D'), (E)** and **(E')** represent the chemical shifts in terms of $^{15}N$ frequency of 800 MHz spectrometer (i.e. 1 ppm along $F_1$ dimension here is equal to ~80 Hz).

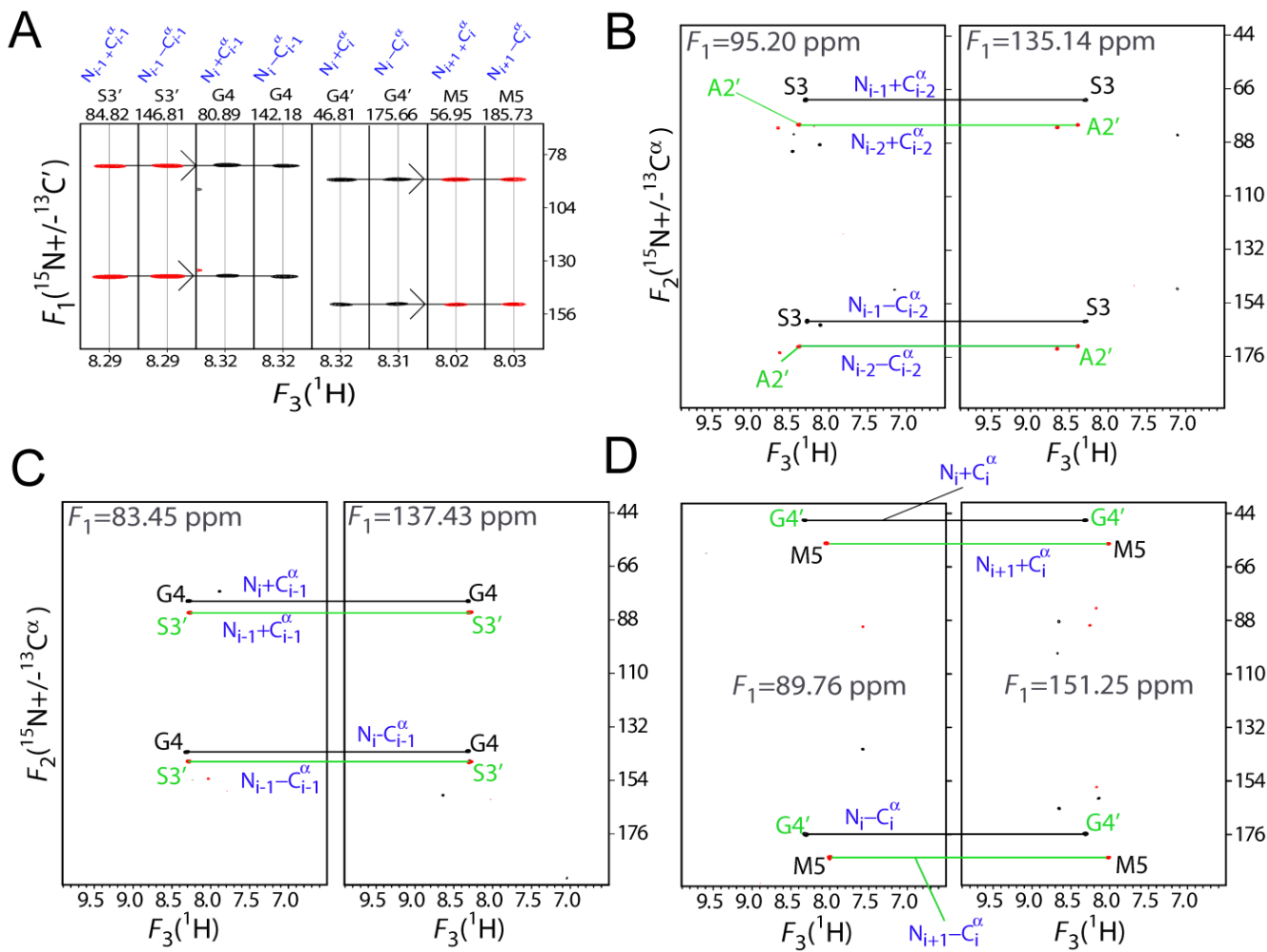

**Figure S4:** An experimental demonstration of L-optimized (5,3)D-h(NCO)-(CAN)H experiment recorded on 15.4 kDa size folded protein referred here as tagged TgADF (see the "Materials and Methods" section of main paper). Like Figure 3 of main paper, it figure shows also the use of correlations observed in the $F_2$–$F_3$ planes of at $N_i^{+/-} (= N_i +/- C'_{i-1})$ chemical shifts of residue $i$, for establishing a sequential ($i \rightarrow i$-1) connectivity between the backbone spin-systems [i.e. $A_i^+/A_i^- \rightarrow B_{i-1}^+/B_{i-1}^-$]. Red and black contours represent positive and negative phase of the peaks, respectively. As evident, a sequential spin-system ($B_{i-1}^+$ and $B_{i-1}^-$) –appearing in both the $F_2$–$F_3$ planes of the spectrum at $N_i^+ / N_i^-$ chemical shifts– can be easily identified because of its opposite peak sign compared to that of the self spin-system $A_i^+$ and $A_i^-$. This feature significantly reduces the search for sequential correlations.

**Table S1:** Acquisition parameters used to record the various spectra on uniformly $^{13}C/^{15}N$ labeled tagged TgADF (at 298 K) and tagged CFP-10 (at 290 K) proteins.

| (5,3)D-h(NCO)(CAN)H | tgADF (folded) Protein | | CFP-10 (unfolded) Protein | |
|---|---|---|---|---|
| | Non-L-Optimized | L-Optimized | Non-L-Optimized | L-Optimized |
| **Complex Data Points ($F_3$ X $F_2$ X $F_1$)** | 1024 X 100 X 40 | 1024 X 100 X 40 | 1024 X 108 X 48 | 1024 X 108 X 48 |
| **Spectral Width and (offset) in ppm** | $F_3(^1H)$ = 12.0 (4.7) $F_2(NC^\alpha)$ =160 (117/70) $F_1(NC')$ =100 (117/186) | $F_3(^1H)$ = 12.0 (4.7) $F_2(NC^\alpha)$ =160 (117/70) $F_1(NC')$ =100 (117/186) | $F_3(^1H)$ = 10.0 (4.7) $F_2(NC^\alpha)$ =140 (119/70) $F_1(NC')$= 90 (119/186) | $F_3(^1H)$ = 10.0 (4.7) $F_2(NC^\alpha)$ =140 (119/70) $F_1(NC')$ =90 (119/186) |
| **No. of scans per FID (Recycle delay)** | 16 (0.8 sec) | 16 (0.3 sec) | 8 (0.8 sec) | 8 (0.3 sec) |
| **Zero filling** | 1024 X 512 X 256 | 1024 X 512 X256 | 1024 X 512 X256 | 1024 X 512 X256 |
| **Total Expt. Time⁺** | ~3 days, 2 hr | ~1 day, 14 hr | ~2 days, 3 hr, 44 min | ~1 day, 3 hr |

**\*Note:** A point to be mentioned here is that the proposed RD experiment require longer time compared to the normal 3D HN(C)N experiment in order to achieve the same resolution. This is because of the fact that one has to acquire more data points along both the co-evolved indirect $F_1$ and $F_2$ dimensions where the linear combination of chemical shifts renders increased spectral width (generally two to three times of the original $^{15}N$ spectral width).